\documentclass{JAC2003}

\usepackage{graphicx}
\usepackage{booktabs}

\begin{document}

\title{
EXPRESSING PROPERTIES OF BPM MEASUREMENT SYSTEM
IN TERMS OF ERROR EMITTANCE AND ERROR TWISS PARAMETERS\vspace{-0.5cm}}

\author{V. Balandin\thanks{vladimir.balandin@desy.de}, W. Decking, N. Golubeva \\
DESY, Hamburg, Germany}

\maketitle

\begin{abstract}
We show that properties of the beam position monitor (BPM) system 
designed for the measurement of transverse and energy beam jitters
can be described in terms of the usual accelerator physics concepts
of emittance, energy spread, dispersions and betatron functions. 
Besides that, using the Courant-Snyder quadratic form as error estimator
we introduce the scalar objective function which can be used as design
criteria of a BPM measurement system with needed properties.
\end{abstract}

\vspace{-0.2cm}

\section{INTRODUCTION}

The determination of variations in the beam position and in the beam energy 
using BPM readings is one of the standard problems
of accelerator physics. If the optical model of the beam line and BPM 
resolutions are known, the typical choice is to let jitter parameters 
be a solution of the weighted linear least squares problem. 
For transversely uncoupled motion this least squares 
problem can be solved analytically,
but the direct usage of the obtained 
solution as a tool for designing a BPM measurement system is not straightforward. 
A better understanding of the nature of the problem is needed.
In this article, following the papers ~\cite{BPM_1, BPM_2, BPM_3},
we show that properties of the BPM measurement system can be described in terms
of the usual accelerator physics concepts of emittance, energy spread, dispersions
and betatron functions. 
Due to space limitation, we consider only
the case of transversely uncoupled
nondispersive beam motion and
inclusion of the energy degree of freedom and multiple examples can be found
in the papers cited above.

\vspace{-0.2cm}

\section{STANDARD LEAST SQUARES SOLUTION}

We assume that the transverse particle motion is uncoupled in linear 
approximation and use the variables $\,\vec{z} = (x, \, p_x)^{\top}\,$ 
for the description of the horizontal beam oscillations.
As transverse jitter parameters 
in the point with the longitudinal position $\,s = r\,$ (reconstruction point)
we mean the difference 
$\,\delta \vec{z}(r) = \vec{z}(r) - \vec{z}_g(r)\,$ 
between parameters of the instantaneous orbit
and parameters of some predetermined ``golden trajectory'' 
$\vec{z}_g = (\bar{x},\, \bar{p}_x)^{\top}$.

Let us assume that we have $n$ BPMs in our beam line placed at positions 
$\,s_1, \ldots, s_n\,$ and they deliver readings
$\vec{b}_{c} = \left(b_1^c, \ldots, b_n^c\right)^{\top}$
for the current trajectory with previously recorded 
observations for the golden orbit being 
$\vec{b}_{g} = \left(b_1^g, \ldots, b_n^g \right)^{\top}$.
Suppose that the difference between these readings 
$\delta \vec{b}_{\varsigma} = \vec{b}_{c} - \vec{b}_{g}$ 
can be represented in the form

\noindent
\begin{eqnarray}
\delta \vec{b}_{\varsigma} =
\left(x(s_1) - \bar{x}(s_1), \ldots,
x(s_n) - \bar{x}(s_n)\right)^{\top} + \vec{\,\varsigma} ,
\label{SEC1_3}
\end{eqnarray}

\noindent
where the random vector 
$ \vec{\,\varsigma} = (\varsigma_1, \ldots, \varsigma_n)^{\top}$ 
has zero mean and positive definite covariance matrix

\noindent
\begin{eqnarray}
V_{\varsigma} \,=\,
\mbox{diag}
\left(\,
\sigma_1^2 ,\, \sigma_2^2 ,\, \ldots ,\, \sigma_n^2\,
\right).
\label{SEC1_17}
\end{eqnarray}

\noindent
As usual, we find an estimate $\delta \vec{z}_{\varsigma} (r)$
for the difference orbit parameters $\delta \vec{z}(r)$
in the presence of BPM reading errors $\vec{\,\varsigma}$ 
by fitting the difference in the BPM data $\delta \vec{b}_{\varsigma}$ 
to the known optical model of the beam line, i.e. by solving
weighted linear least squares problem.
If the phase advance between at least 
two BPMs is not a multiple of $180^{\circ}$, 
then the result of this fit is unique 
and is given by the formula

\noindent
\begin{eqnarray}
\delta \vec{z}_{\varsigma}(r)  = 
\left(M^{\top}(r) V_{\varsigma}^{-1} M(r) \right)^{-1}
M^{\top}(r) \, V_{\varsigma}^{-1} \cdot \delta \vec{\,b}_{\varsigma}.
\label{SEC1_15}
\end{eqnarray}

\noindent
The calculation of the covariance matrix of the reconstruction errors 
is also standard and gives the following result

\noindent
\begin{eqnarray}
V_z(r) \, \stackrel{\rm def}{=} \,
{\cal V} \left(\, \delta \vec{z}_{\varsigma}(r) \, \right)
\,=\,
\left(M^{\top}(r) \, V_{\varsigma}^{-1} \, M(r) \right)^{-1}.
\label{SEC1_16}
\end{eqnarray}

\noindent
Here 

\vspace{-0.2cm}
\noindent
\begin{eqnarray}
M \,=\,
\left(
\begin{array}{cc}
a_{11}(r,\,s_1) & a_{12}(r,\,s_1)\\
\vdots & \vdots\\
a_{11}(r,\,s_n) & a_{12}(r,\,s_n)
\end{array}
\right),
\label{a2}
\end{eqnarray}

\noindent
and $\,a_{11}(r_1,\,r_2),\,a_{12}(r_1,\,r_2)\,$ are the elements
of a two by two symplectic matrix $\,A(r_1, \, r_2)\,$ 
which transfers particle coordinates from 
the point with the longitudinal position  $\,s = r_1\,$ to the point
with the position  $\,s = r_2$. 

For the considered one-dimensional case the matrix inversion
in the right hand side of the formula (\ref{SEC1_16}) can be 
done analytically and the elements of the error covariance matrix $V_z(r)$
can be obtained in the explicit form as follows

\noindent
\begin{eqnarray}
\left(V_z(r)\right)_{1,1} =
\frac{ 1 }{ \Delta }
\sum \limits_{ m = 1 }^{ n } \left( \frac{a_{12}(r, s_m)}{\sigma_m} \right)^2,   
\label{SEC1_18_1_1}
\end{eqnarray}

\vspace{-0.2cm}

\noindent
\begin{eqnarray}
\left(V_z(r)\right)_{1,2} =
\left(V_z(r)\right)_{2,1} =
\nonumber
\end{eqnarray}

\vspace{-0.2cm}

\noindent
\begin{eqnarray}
-\frac{ 1 }{ \Delta }
\sum \limits_{ m = 1 }^{ n } \left( \frac{a_{11}(r, s_m)}{\sigma_m} \right) 
\left( \frac{a_{12}(r, s_m)}{\sigma_m} \right),
\label{SEC1_18_1_2}
\end{eqnarray}

\vspace{-0.2cm}

\noindent
\begin{eqnarray}
\left(V_z(r)\right)_{2,2} =
\frac{ 1 }{ \Delta }
\sum \limits_{ m = 1 }^{ n } \left( \frac{a_{11}(r, s_m)}{\sigma_m} \right)^2,   
\label{SEC1_18_2_2}
\end{eqnarray}

\noindent
where

\noindent
\begin{eqnarray}
\nonumber
\Delta = 
\frac{1}{2}
\sum \limits_{k, m = 1 }^{ n }
\;\;\;\;\;\;\;\;\;\;\;\;\;\;\;\;\;\;\;\;\;\;\;\;\;\;\;\;\\
\left( \frac{a_{11}(r, s_k) a_{12}(r, s_m) 
- a_{11}(r, s_m) a_{12}(r, s_k)}{\sigma_k \sigma_m} \right)^2.
\label{SEC1_19}
\end{eqnarray}

\noindent
In theory, the formulas (\ref{SEC1_18_1_1})-(\ref{SEC1_19}) contain 
all information which one has to know in order to be able to
design a BPM measurement system with needed properties or in order
to be able to compare expected performance of two different
measurement systems. 
In practice, unfortunately, the simple knowledge of formulas
(\ref{SEC1_18_1_1})-(\ref{SEC1_19})
is quite far from being sufficient for these purposes.
Let us assume, for example, that we want to compare resolutions
of two BPM systems which are supposed to be used for trajectory
jitter determination and are installed in two different beam lines.
For this purpose we need, at least, to have a criteria how to compare
two covariance matrices and to know how to chose the reconstruction points
(own for each measurement system) in which such comparison has
to be done.
Does all that looks to be fairly
straightforward?

\section{ERROR TWISS PARAMETERS AND COURANT-SNYDER QUADRATIC FORM 
AS ERROR ESTIMATOR}

An important step in solving the problems 
marked at the end of the previous section
was made in ~\cite{BPM_1, BPM_2}, where 
dynamics was introduced into this problem which in the beginning 
seemed to be static. When one changes the position of the reconstruction 
point, the estimate of the jitter parameters propagates along the beam
line exactly as a particle trajectory 
and it becomes possible (for every fixed
jitter values) to consider a virtual beam consisting of virtual particles 
obtained as a result of application of least squares reconstruction procedure 
to ``all possible values'' of BPM reading errors. The dynamics of the centroid 
of this beam coincides with the dynamics of the true difference orbit 
and the covariance matrix of the jitter reconstruction errors can be treated
as the matrix of the second central moments of this virtual beam
distribution and satisfies the usual transport equation

\noindent
\begin{eqnarray}
V_z(r_2) \;=\; A( r_1 , \, r_2 ) \, V_z(r_1) \,
A^{\top}( r_1 , \, r_2 ) .
\label{bdp_3}
\end{eqnarray}

\noindent
Consequently, for the description of the propagation of
the reconstruction errors along the beam line,
one can use the accelerator physics notations
and represent the error covariance matrix in the familiar form

\noindent
\begin{eqnarray}
V_z(r) =
\epsilon_{\varsigma} 
\, 
\left(
\begin{array}{rrr}
  \beta_{\varsigma}(r) & -\alpha_{\varsigma}(r)\\
-\alpha_{\varsigma}(r) &  \gamma_{\varsigma}(r)
\end{array}
\right),
\label{TwCVM_1}
\end{eqnarray}

\noindent
where $\beta_{\varsigma}(r)$ and $\alpha_{\varsigma}(r)$
are the error Twiss parameters and

\noindent
\begin{eqnarray}
\epsilon_{\varsigma} \,=\, \sqrt{\,\det V_z(r)\,} \,=\, 1 / \sqrt{\Delta}
\label{SEC2_5}
\end{eqnarray}

\noindent
is the invariant error emittance.

Note that the error Twiss parameters 
can also be found as solution of the minimization
problem

\noindent
\begin{eqnarray}
\min_{\beta(r), \, \alpha(r)}
\;\;\sum \limits_{ m = 1 }^{ n } \frac{\beta(s_m)}{\sigma_m^2} . 
\label{SEC2_11}
\end{eqnarray}

\noindent
Under the assumption that the phase advance between at least
two BPMs is not a multiple of $180^{\circ}$, 
the solution of the problem (\ref{SEC2_11}) is unique,
the minimum is reached at the error Twiss parameters and
is equal to $\,2 / \epsilon_{\varsigma}$.

Parametrization (\ref{TwCVM_1}) is an essential step ahead in 
understanding of the structure of the matrix $V_z(r)$
in comparison with the formulas (\ref{SEC1_18_1_1})-(\ref{SEC1_19}).
It introduces such important characteristic of BPM measurement
system as error emittance and shows that 
balance between coordinate and momentum reconstruction errors 
in the point of interest is defined 
by the values of error Twiss parameters at this location. 
Nevertheless, it still does not give a single property
to compare two different BPM systems.
Fortunately, the beam dynamical point of view on the BPM measurement system
naturally suggests us that in order to obtain the needed criteria 
we may simply use the Courant-Snyder 
quadratic form as an error estimator.

Let $\,\beta_0, \, \alpha_0, \, \gamma_0\,$ be the 
design Twiss parameters and

\noindent
\begin{eqnarray}
I_x (r, \, \vec{z}\,) = 
\gamma_0(r) \,x^{2} + 2 \alpha_0(r) \,x p_x  + \beta_0(r) \,p_x^{2} 
\label{IFB_1}
\end{eqnarray}

\noindent
the corresponding Courant-Snyder quadratic form. 
Using this quadratic form
we introduce the random variable

\noindent
\begin{eqnarray}
I_x^{\varsigma} \,=\, I_x ( r, \, \delta \vec{z}_{\varsigma}(r) -
\delta \vec{z}(r)) .
\label{SEC_GD_1}
\end{eqnarray}

\noindent
The mean value of this random variable is equal

\noindent
\begin{eqnarray}
\big< \, I_x^{\varsigma} \, \big> 
\,=\, 2 \,\epsilon_{\varsigma} \, m_p(\beta_{\varsigma},\,\beta_0) ,
\label{App1}
\end{eqnarray}

\noindent
where

\noindent
\begin{eqnarray}
m_p(\beta_{\varsigma},\,\beta_0) =
(\beta_{\varsigma}\gamma_0 - 2 \alpha_{\varsigma}\alpha_0 + \gamma_{\varsigma}\beta_0) \,/\, 2.
\label{AppMP}
\end{eqnarray}

\noindent
The right hand side in (\ref{App1}) does not depend on the position of the reconstruction 
point (is a number), characterizes the resolution of the BPM system 
not in some absolute units but in the relative units of beam sigmas
and, therefore, allows to compare properties of 
two completely different BPM systems installed in two different beam lines
and also can be used as scalar valued (not matrix valued) design criteria.

\section{Two BPM Case}

Let us consider two BPMs separated in the beam line by a transfer matrix
$A(s_1, s_2)$ with $a_{12} \neq 0$
and assume that these BPMs deliver uncorrelated readings
with rms resolutions $\sigma_1$ and $\sigma_2$ respectively. 
Often, when one works on optimization of the optics of two
BPM system, one speaks about the desire to have 
the large beta functions at the BPM locations and
the phase advance being equal or enough close to $90^{\circ}$ .
And that is completely right, if one will interpret
this desire as a way to increase absolute value
of the $\,a_{12}\,$ coefficient, because
the error emittance of the two BPM measurement system is 
inversely proportional to it

\noindent
\begin{eqnarray}
\epsilon_{\varsigma} \;=\;
(\sigma_1 \, \sigma_2) \,/\,\left| a_{12}\right|.
\label{SEC_TwoBPM_1}
\end{eqnarray}

\noindent
Nevertheless, because due to formula (\ref{App1})
the figure of merit for the quality of BPM system is not
the error emittance alone, but the product of the error emittance and the mismatch 
between the error and the design Twiss parameters (large mismatch
can spoil the properties of the measurement system
even for the case when the error emittance is small), 
one has to take additional care and compare design betatron
functions with the error betatron functions which are given below

\noindent
\begin{eqnarray}
\beta_{\varsigma}(s_1) =
\frac{\sigma_1}{\sigma_2} \left| a_{12}\right|,
\hspace{0.3cm}
\alpha_{\varsigma}(s_1) =\;\;\,
\frac{\sigma_1}{\sigma_2} \mbox{sign}\left(a_{12}\right) a_{11},\\
\beta_{\varsigma}(s_2) =
\frac{\sigma_2}{\sigma_1} \left| a_{12}\right|,
\hspace{0.3cm}
\alpha_{\varsigma}(s_2) =
- \frac{\sigma_2}{\sigma_1} \mbox{sign}\left(a_{12}\right) a_{22}.
\label{SEC_TwoBPM_3}
\end{eqnarray}

Let us note that though the error Twiss parameters
depend on the ratio of BPM resolutions, the error phase advance 
(phase advance defined by $\beta_{\varsigma}$) 
is independent from this ratio and is
always equal to an odd multiple of $90^{\circ}$.

\section{Periodic Measurement System}

In this section we consider a measurement system constructed from $n$ 
identical cells assuming that we have one BPM per cell (identically positioned 
in all cells) and that the cell transfer matrix allows periodic beam 
transport with phase advance $\,\mu_p\,$ being not a multiple 
of $\,180^{\circ}$. Additionally, we assume that all BPMs have 
the same rms resolution $\,\sigma_{bpm}$. 
In this situation the formula for the
error emittance is rather simple and is given by the expression

\noindent
\begin{eqnarray}
\epsilon_{\varsigma} \;=\;
\frac{2 \sigma_{bpm}^2}{ n \beta_p (s_1)} \cdot
m_p(\beta_{\varsigma},\,\beta_p)
\,,
\label{SEC4_1}
\end{eqnarray}

\noindent
where $\,\beta_p(s_1)$ is the value of the periodic betatron function 
at the BPM locations and

\noindent
\begin{eqnarray}
m_p(\beta_{\varsigma},\,\beta_p) \;=\;
\left(
\,1 \,-\,  \left(\frac{1}{n}\cdot
\frac{\sin(n \mu_p)}{\sin(\mu_p)}\right)^2\,
\right)^{-\frac{1}{2}}
\label{SEC4_2}
\end{eqnarray}

\noindent
is the mismatch between the error and the periodic betatron functions
(even so we do not assume, in general, periodic betatron functions being the design
betatron functions matched to our beam line).

There is a rather widespread opinion that a periodic measurement system
reaches an optimal performance 
when its design Twiss parameters are cell periodic and
the cell phase advance is a multiple of $180^{\circ}$ divided by $n$. 
Is that really so? 
To answer this question let us take the cell periodic Twiss parameters as
design Twiss parameters and write

\noindent
\begin{eqnarray}
\left< I_x \right> \;=\; 
\frac{4 \sigma_{bpm}^2}{ n \beta_p (s_1,\,\mu_p)} \cdot
m_p^2(\beta_{\varsigma},\,\beta_p)
\,.
\label{SEC4_3_1_0}
\end{eqnarray}

\noindent
Looking at the formula (\ref{SEC4_2})
one sees that the choice of $\mu_p$ 
such that $\sin(n \mu_p) = 0$
makes the error and the periodic Twiss parameters 
equal and brings the second multiplier in the right hand side
of the formula (\ref{SEC4_3_1_0}) to the minimal possible value.
But, in general, it does not guarantee that the product of the two
multipliers in (\ref{SEC4_3_1_0}) is also minimized.
So the answer is not or, more exactly, not necessary.

To be more specific, let us consider a 
thin lens FODO cell of the length $\,L\,$ 
as a basic unit of our periodic system and let us also assume that 
the BPM is placed in the ``center" of
the focusing lens.
In this situation 

\noindent
\begin{eqnarray}
\left< I_x \right> \;=\; 2 \epsilon_{\varsigma}
m_p\left(\beta_{\varsigma}, \, \beta_p\right) \;=\;
\frac{4 \sigma_{bpm}^2}{n L} \cdot \Psi_n \left( \mu_p \right)\,,
\label{SEC4_3_1}
\end{eqnarray}

\noindent
where

\noindent
\begin{eqnarray}
\Psi_n \left( \mu_p \right) \;=\;\Psi_{\infty} \left( \mu_p \right)
\cdot 
m_p^2\left(\beta_{\varsigma},  \,\beta_p\right),\;\;\\
\Psi_{\infty} \left( \mu_p \right) \;=\;
\sin(\mu_p) \,/\, (1 + \sin(\mu_p / 2)).
\label{SEC4_3_3}
\end{eqnarray}

\noindent
The functions $\Psi_n$ for $n = 2, 3, 4, 5$
are plotted in figure 1 
together with their values in the points

\noindent
\begin{eqnarray}
\mu_p \;=\; k \cdot (180^{\circ}\,/\,n),
\hspace{0.3cm}
k = 1, \ldots , n - 1,
\label{SEC4_3_4}
\end{eqnarray}

\noindent
shown as small circles at the corresponding curves.
One sees that for all $n$ the optimal
performance of our measurement system 
(minimum of $\Psi_n$)
is reached for the
phase advance which is different from
the multiples of $180^{\circ}\,/\,n$.

\begin{figure}[t]
    \centering
    \includegraphics*[width=80mm]{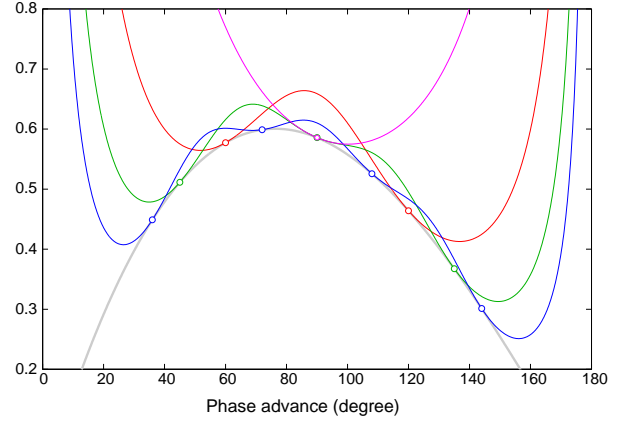}
    \vspace{-0.2cm}
    \caption{Functions $\,\Psi_n \left( \mu_p \right)\,$ shown for
    $\,n = 2, 3, 4, 5\,$ (magenta, red, green and blue curves respectively).
    The gray curve shows function $\,\Psi_{\infty} \left( \mu_p \right)\,$.}
    \vspace{-0.3cm}
    \label{fig_Psi1}
\end{figure}

\end{document}